\documentclass[conference]{IEEEtran}
\usepackage{amsmath}
\usepackage{epsfig}
\usepackage{multirow,multicol}
\usepackage{url}
\usepackage{subfigure}
\usepackage{balance}

\long\def\symbolfootnote[#1]#2{\begingroup
\def\thefootnote{\fnsymbol{footnote}}
\footnote[#1]{#2}\endgroup}

\psfull

\allowdisplaybreaks[1]

\begin{document}

\title{Age-of-Information for Computation-Intensive Messages  in Mobile Edge Computing
}
\author{\IEEEauthorblockN{Qiaobin~Kuang\IEEEauthorrefmark{1}, Jie~Gong\IEEEauthorrefmark{2}, Xiang~Chen\IEEEauthorrefmark{1}, and Xiao Ma \IEEEauthorrefmark{2}
\IEEEauthorblockA{\IEEEauthorrefmark{1}School of Electronics and Information Technology, Sun Yat-sen University, Guangzhou 510006, China}
\IEEEauthorblockA{\IEEEauthorrefmark{2}School of Data and Computer Science, Sun Yat-sen University, Guangzhou 510006, China}
Email: kuangqb@mail2.sysu.edu.cn,~gongj26@mail.sysu.edu.cn,~chenxiang@mail.sysu.edu.cn,~maxiao@mail.sysu.edu.cn
}
}
\setlength\abovedisplayskip{3pt}
\setlength\belowdisplayskip{3pt}
\maketitle
%\and

\maketitle

\begin{abstract}
Age-of-information (AoI) is a novel metric that measures the freshness of information in status update scenarios. It is essential for real-time applications to transmit status update packets to the destination node as timely as possible. However, for some applications, status information embedded in the packets is not revealed until complicated data processing, which is computational expensive and time consuming. As mobile edge server has sufficient computational resource and is placed close to users, mobile edge computing (MEC) is expected to reduce age for computation-intensive messages. In this paper, we study the AoI for computation-intensive data in MEC, and consider two schemes: local computing by user itself and remote computing at MEC server. The two computing models are unified into a two-node tandem queuing model. Zero-wait policy is adopted, i.e., a new message is generated once the previous one leaves the first node. We consider exponentially distributed service time and infinite queue size, and hence, the second node can be seen as a First-Come-First-Served (FCFS) M/M/1 system. Closed-form average AoI is derived for the two computing schemes. The region where remote computing outperforms local computing is characterized. Simulation results show that the remote computing is greatly superior to the local computing when the remote computing rate is large enough, and that there exists an optimal transmission rate so that remote computing is better than local computing for a largest range.
% And there exists a optimal packet size to minimize the AoI for remote computing.
\end{abstract}

%\begin{IEEEkeywords}
%Age-of-information, Mobile edge computing; computation-intensive.
%\end{IEEEkeywords}

\section{Introduction}
In recent years, real-time applications such as weather monitoring, stocks forecast, and social updates have drawn great attention. In these applications, maintaining the freshness of information is important for accurate monitoring. For accurate status acquisition in these applications, it is essential to maintain the freshness of data, which is measured by age-of-information (AoI) \cite{Kaul2012Real}, also referred to as age, defined as the time elapsed since the generation of the latest delivered update. Conventional researches mainly focus on the impact of data transmission and queuing on AoI. However, in applications such as autonomous driving, an update like an image needs not only to be transmitted to the controller, but also to be processed before the useful information embedded in the image is exposed, which could be computational expensive and time consuming due to limited computational capacity of local processors. Mobile edge computing (MEC) can not only  provide sufficient computing resource near the user, but also reduce the response time compared with the centralized cloud \cite{Mao2017A}. Thus, AoI for computation-intensive messages can be reduced by adopting MEC.

AoI was initially proposed in \cite{Kaul2012Real},\cite{Kaul2011Minimizing} to measure the freshness of information at the destination node. Since then, there have been numerous works about AoI. Ref. \cite{Roy2012Real} studied multiple sources status updating at interested recipients. The authors in \cite{Sun2017Update} found that the zero-wait policy is far from the optimum in some cases. The works in \cite{Kadota2016Minimizing,Hsu2017Age} focus on broadcast wireless networks. In particular, a transmission scheduling policy was proposed in \cite{Kadota2016Minimizing} to optimize AoI in a broadcast wireless network over unreliable channels, and  AoI in a wireless broadcast network where only one user can be served at a time was studied in \cite{Hsu2017Age}. For  multi-hop wireless networks, minimizing AoI in multi-hop interference free networks with a single information flow was firstly considered in \cite{Ahmed2017The}, and  AoI in multi-hop wireless networks with multi source-destination pairs and general interference constrains was analyzed in \cite{Talak2017Minimizing}. Recently, some works are devoted to developing new tools for AoI analysis in networks. In particular, in \cite{Roy2018Age}, an explicit calculation of the average age was calculated  over a multi-hop network of preemptive servers by using a stochastic hybrid system (SHS). And in \cite{Yates2018Age}, the authors applied SHS to analyze the temporal convergence of higher order AoI moments, and enable the moment generation function to characterize the stationary distribution of an AoI process in multi-hop networks.

As seen in the existing works, AoI is mainly influenced by the packet generation frequency as well as the delay caused by data transmission and queuing. However, for computation-intensive messages, data processing delay is not negligible. For application such as autonomous driving, online games and augmented reality, a large amount of image processing, voice recognition is performed to identify the real status. Among the limited research efforts, Ref.\cite{Alabbasi2018Joint} jointly considered the computation and information freshness for vehicular networks and proposed novel scheduling strategies for both computing and network stages. In particular, each vehicle updates the data on the cloud at certain rate, and request computation for some tasks at certain rate. The model is mainly builded from the perspective of the cloud server, and analyzing the computation phase at the server and the networking phase of transferring the results from the server to the user. In this paper, we focus on the computing process that can be executed either local server or MEC server, as well as the transmission from the source node which generate the computation-intensive messages to the MEC server if the messages are processed at MEC server. To the best of our knowledge, AoI in MEC has not been studied.

In this paper, we study the AoI performance for computation-intensive messages in MEC for two schemes. One is local computing, and the other is remote computing. Zero-wait policy is applied to computing process in local computing and transmission process in remote computing at MEC server. The computing process in remote computing follows first-come-first-served (FCFS) principle. The two schemes can be unified as a two-node tandem queuing model. In remote computing, assuming that both transmission time and computing time follow exponential distribution, and the queue size is infinite. Then, the computing process can be viewed as an FCFS M/M/1 system. In this model, we obtain the average AoI for local computing and remote computing. Based on the AoI expressions, we characterized the region where remote computing outperforms local computing. The impact of the packet size, the required number of CPU cycles, the data rate and the computing capacity of MEC server for data processing on the average AoI is studied by numerical results. It is shown that the AoI in remote computing becomes small as the required number of CPU cycles decreases and the computing capacity of MEC server increases. Considering only the effect of the packet size or data rate on the average age in remote computing, there exists an optimal packet size and data rate to minimize the average AoI. For the local computing, the average AoI becomes small as the required number of  CPU cycles decreases but keeps constant while changing the packet size and the data rate. It is also found that remote computing outperforms local computing at lager required number of CPU cycles or computing capacity of MEC server, as well as appropriate  packet size and data rate.

\section{Model and Formulation}
Consider a status monitoring and control system for computation-intensive messages as shown in Fig. \ref{fig:subfig}. System status is generated by the source, and processed by either local or remote computing. Then, the processed signal is sent to the destination node. For the accuracy of control, the processed status should be as fresh as possible. In the following, we will describe the local and remote computing in detail.
\subsection{Local computing and remote computing model}
\begin{figure}
  \centering
  \subfigure[Local computing]{
    \label{fig:subfig:a} %% label for first subfigure
    \includegraphics[width=3.2in]{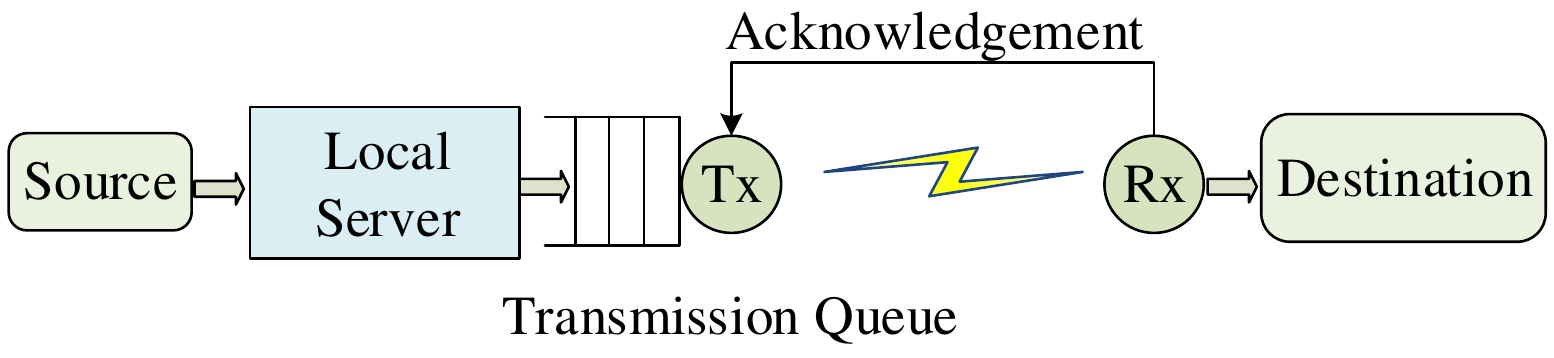}
  }
  \subfigure[Remote computing]{
    \label{fig:subfig:b} %% label for second subfigure
    \includegraphics[width=3.4in]{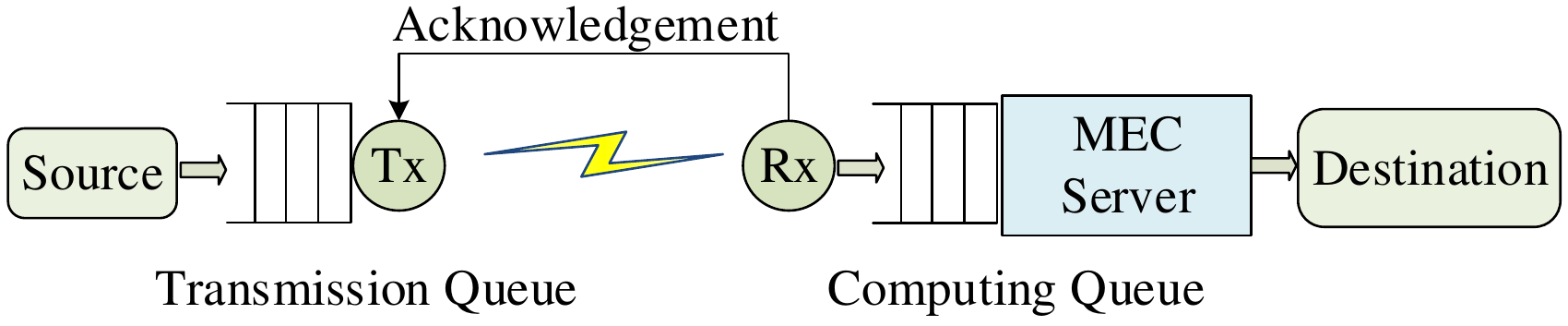}
  }
  \caption{System models}
  \label{fig:subfig} %% label for entire figure
\end{figure}
As both MEC server and user have the capacity to compute, we compare two schemes in this paper. One is computing the computation-intensive data locally, and sending the result to the destination node as shown in Fig. \ref{fig:subfig:a}. This scheme is termed as \emph{local computing}. The other is transmitting the computation-intensive packet to the MEC server to compute remotely as shown in Fig. \ref{fig:subfig:b}, abbreviated as \emph{remote computing}. Zero-wait policy \cite{Sun2017Update} is adopted in both local computing and remote computing. In particular, for local computing, a new status update packet is generated when the last packet is completely computed by itself. In remote computing, a new status update packet is generated and starts transmission when the last status update packet is delivered to the receiver. For local computing, the queuing delay is completely eliminated by zero-wait policy. Since the size of the control signal after computing is much smaller than the size of the original packet, the time for the control signal to be transmitted to the destination node can be ignored relative to the time for  computing. For remote computing, the queuing delay for transmission is zero. And the MEC server applies FCFS principle to the delivered status update packets. Thus, some packets have to wait in the queue to be processed due to the randomness of computing time.
\subsection{Unified model}
\begin{figure}
\centering
    \includegraphics[width=3.4in]{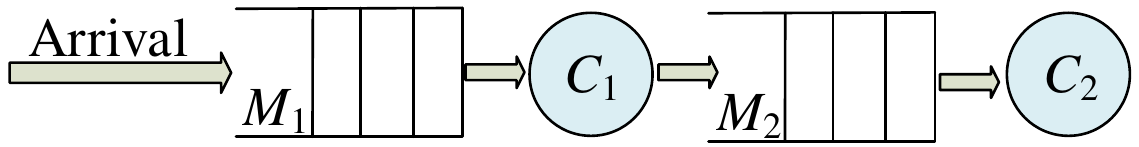}
\caption{A unified two-node tandem  queuing model} \label{fig:jointfig}
\end{figure}

The two schemes can be unified as a two-node tandem model. For remote computing, as shown in Fig.  \ref{fig:jointfig}, the server $C_1$ refers to transmission channel, $C_2$ refers to the MEC  server, $M_1$ refers to the transmission queue and $M_2$ refers to the computing queue. While  local computing  can be viewed as a special case with service rate of $C_2$ going infinity, and $C_1$ refers to local computing server, $M_1$ refers to the computing queue which is empty due to the zero-wait policy, and $M_2$ is empty as well. We will derive AoI based on this unified model.
\begin{figure}
\centering
    \includegraphics[width=3.5in]{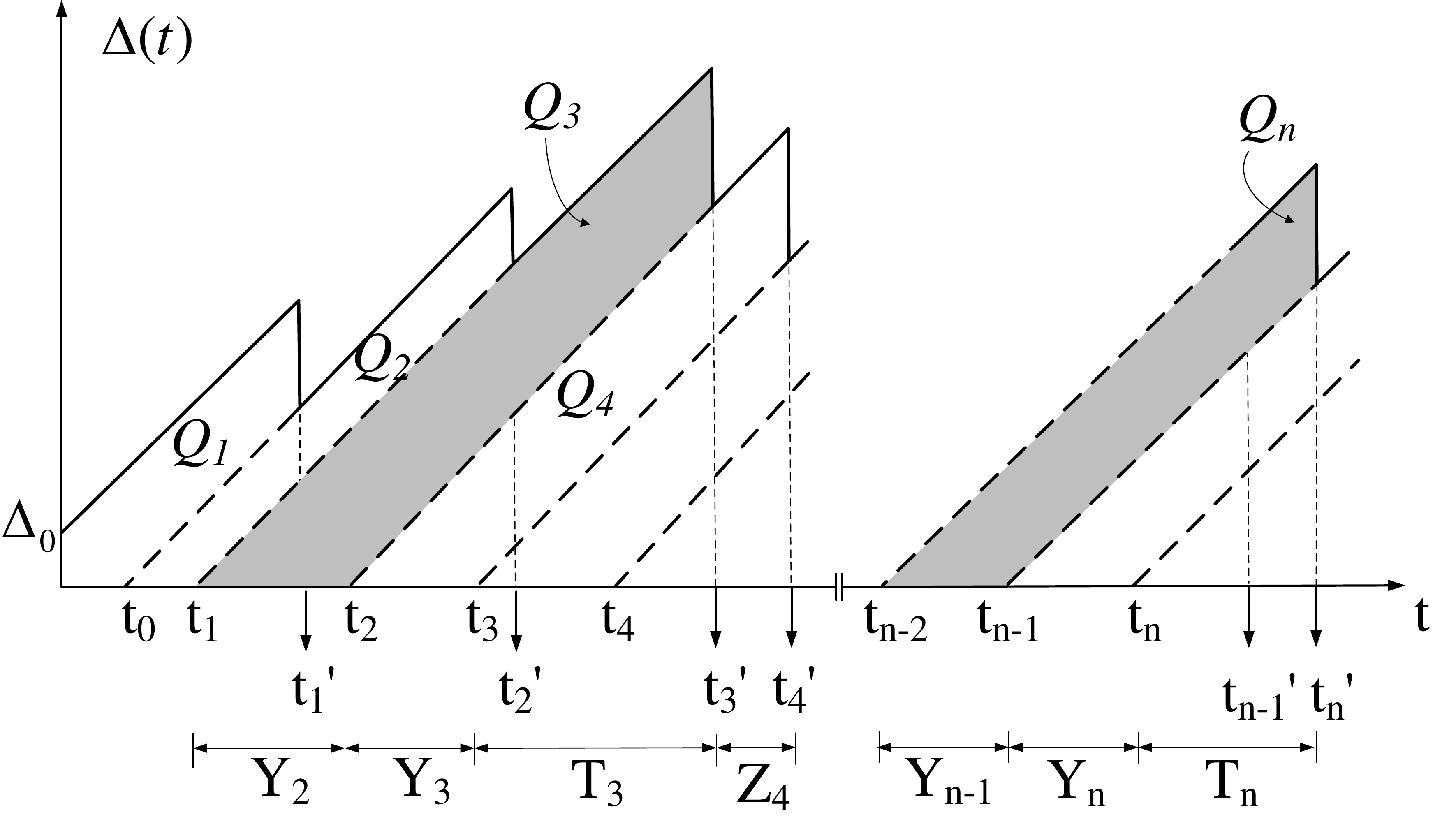}
\caption{Evolution of the age-of-information $\Delta(t)$} \label{fig:evolution}
\end{figure}
The latest processed packet at time $t$ is time-stamped $u(t)$ representing its generation time. The AoI of the processed status in the destination node at time $t$ is defined as the random process
\begin{equation}\label{equ:}
\Delta(t) = t-u(t).
\end{equation}
 The evolution of the AoI $\Delta(t)$ at the destination under FCFS queuing is shown in Fig. \ref{fig:evolution}. Without loss of generality, we start from observing $t = 0$ and the queue is empty with $ \Delta(0) = \Delta_0 $. As shown in Fig. \ref{fig:evolution}, the $i$-th status update packet arrives at $M_2$ at time instant $t_i$. According to zero-wait policy, $t_i$ is also the service starting time instant in $C_1$ for the $(i+1)$-th status update packet with service rate $\mu_1$. Denote $t_i^{'}$ as the service termination time instant of the $i$-th status update packet in $C_2$ with service rate $\mu_2$. The age at the destination increases linearly without service termination in $C_2$ and rapidly reduce to a smaller value otherwise, i.e., updating the processed status at the destination. Both service times are assumed independent and identically distributed (iid) with exponential distribution.

The average age of the processed status packet is the function $\Delta(t)$ in Fig.  \ref{fig:evolution} normalized by the time interval observed. During the interval $(0, \tau )$, the average AoI is \cite{Kaul2012Real}
\begin{equation}\label{equ:}
\Delta_\tau = \frac{1}{\tau}\int_0^{\tau}\Delta(t)dt.
\end{equation}
For simplicity of exposition, we set the length of the observation interval $\tau = t_{n}^{'}$. As depicted in Fig.  \ref{fig:evolution}, the average age can be represented as
\begin{equation}\label{equ:defi}
\Delta_\tau = \frac{\sum_{i =1}^{n}Q_i+(Y_{n}+T_{n})^2/2}{\tau}.
\end{equation}
From Fig.  \ref{fig:evolution}, we know that $Q_i (i> 1)$ is an isosceles trapezoid, which can be derived from two isosceles triangles, i.e.,
\begin{equation}\label{equ:Q}
\begin{split}
Q_i &= \frac{1}{2}(Y_{i-1}+Y_{i}+T_{i})^2-\frac{1}{2}(Y_{i}+T_{i})^2\\
&= T_{i}Y_{i-1}+Y_{i}Y_{i-1}+\frac{1}{2}Y_{i-1}^2,
\end{split}
\end{equation}
where $Y_i=t_i-t_{i-1}$ denotes the inter-arrival time between the $(i-1)$-th packet and the $i$-th packet at $M_2$, which is equivalent to the service time of packet $i$ in $C_1$, and $T_{i}=t_{i}^{'}-t_{i}$ is the elapsed time between the service termination time instant in $C_2$ and the arrival time instant at $M_2$ for the $i$-th status update packet. The average AoI can be rewritten as
\begin{equation}\label{equ:de_inf}
\Delta_\tau = \frac{\tilde{Q}}{\tau}+\frac{n-1}{\tau}\frac{1}{n-1}\sum_{i =2}^{n}Q_i,
\end{equation}
where $\tilde{Q}=Q_1+(Y_{n}+T_{n})^2/2$. Note that $\tilde{Q}$ is finite,
as $\tau\rightarrow \infty$, the first term in (\ref{equ:de_inf}) will be zero. The term $\frac{n-1}{\tau}$ will be the steady service rate of $C_1$ as $\tau$ increases. Thus, we can get the following equation
\begin{equation}\label{equ:de_mu1}
\mu_1= \lim_{\tau\rightarrow\infty}\frac{n}{\tau}.
\end{equation}
 Substituting (\ref{equ:Q}) into (\ref{equ:de_inf}), and let  $\tau$ goes to infinity, the average status update age can be expressed as \cite{Kaul2012Real}
\begin{equation}\label{equ:average_1}
\begin{split}
\bar{\Delta}&= \lim_{\tau\rightarrow \infty}\Delta_\tau =\mu_1E[Q_i]\\
&=\mu_1\left(E[T_{i}Y_{i-1}]+E[Y_{i}Y_{i-1}]+\frac{1}{2}E[Y_{i-1}^2]\right),
\end{split}
\end{equation}
where $E(\cdot)$ is the expectation operator. In order to obtain the average AoI, we need to calculate three expectations in the above equation, which is detailed in the next section.
\section{Average AoI}
In this section, we firstly present the main result of this paper, i.e., the closed-form expression of average AoI for two-node tandem queuing model and make comparison with \cite{Kaul2012Real}, and then describe the detailed calculation. Finally, the result is applied to both local computing and remote computing.
\subsection{Main Result}
%\begin{thm}\label{thm1}
The average age is expressed as the following equation
\begin{equation}\label{equ:average_up}
\bar{\Delta} = \frac{1}{\mu_2}\left(\frac{\rho(2\rho^2-\rho+1)}{(1+\rho)(1-\rho)}+\frac{2}{\rho}+1\right),
\end{equation}
where $\rho =\mu_1/\mu_2$ denotes the server utilization of $C_2$.

It is interesting to compare with the result in \cite{Kaul2012Real}. As shown in Fig. \ref{fig:evolution}, denote $Z_i$ as the inter-departure time from $C_2$ between the $(i-1)$-th packet and the $i$-th packet. Thus, the area $Q_i (i>1)$ can be re-expressed as
\begin{equation}\label{}
\begin{split}
Q_i &= \frac{1}{2}(Y_{i-1}+T_{i-1}+Z_{i})^2-\frac{1}{2}(T_{i-1}+Z_{i})^2\\
&= T_{i-1}Y_{i-1}+\frac{1}{2}Y_{i-1}^2+Z_{i}Y_{i-1}.
\end{split}
\end{equation}
Notice that the expression of $Q_i$ in the M/M/1 FCFS system (equation (4) in [1]) is $Q_i=T_{i}Y_{i}+\frac{1}{2}Y_{i}^2$. The inter-arrival time $Y_i$ are iid exponentials. Moreover, the system will reach a steady state, i.e., $T_{i}=T_{i-1}$, more details can be seen in the following subsection. Thus, the following equations can be obtained
\begin{equation}\label{}
T_{i-1}Y_{i-1}=T_{i}Y_{i},
\end{equation}
\begin{equation}\label{}
\frac{1}{2}Y_{i-1}^2=\frac{1}{2}Y_{i}^2.
\end{equation}
 Therefore, in this paper, the area $Q_i$ is the area for M/M/1 FCFS system plus that of the parallelogram, that is $Z_{i}Y_{i-1}$. Comparing the results in this paper and \cite{Kaul2012Real}, the expectation of the term $Z_{i}Y_{i-1}$ can be derived as
\begin{equation}\label{equ:ex_comp}
E[Z_{i}Y_{i-1}]=\frac{1}{\mu_1^2}+\frac{1-\rho}{\mu_2^2(1+\rho)}.
\end{equation}
 It is easy to verify that $E[Z_i] = E[Y_{i-1}] = 1/\mu_1$. Therefore, the inter-departure time $Z_i$ from $C_2$ and the inter-arrival time $Y_{i-1}$ at $M_2$ are dependent. The reason is that $Y_{i-1}$ is related to the system time $T_{i}$, which then has effect on $Z_i$. Although we can derive the average AoI by summing the average AoI in \cite{Kaul2012Real} and the additional area, equation (\ref{equ:ex_comp}) is difficult to be obtained directly. In the following subsection, we will introduce a more convenient process to calculate the average AoI based on (\ref{equ:Q}).

\subsection{Calculation of Average AoI}
Notice that the arrival process of $M_2$ is equivalent to the departure process of $C_1$, which is a Poisson process due to zero-wait policy. Thus, $M_2$ and $C_2$ forms an FCFS $M/M/1$ system. Therefore, both the inter-arrival time $Y_i$ and the service time are iid exponentials with $E[Y_i] = 1/\mu_1$ and average service time $1/\mu_2$, respectively. As $Y_{i-1}$ and $Y_i$ are independent, we have
\begin{equation}\label{equ:exp_y}
E[Y_{i}Y_{i-1}]=(E[Y_i])^2 =1/\mu_1^2,
\end{equation}
\begin{equation}\label{equ:exp_y_sq}
 E[Y_{i-1}^2]=2/\mu_1^2.
 \end{equation}
 Then we calculate $E[T_{i}Y_{i-1}]$ in detail. For status update $i$, $T_i$ also represents the system time in queuing theory, which consists of waiting time and service time, i.e.,
\begin{equation}\label{equ:sys_time}
T_{i} = W_i+S_i,
\end{equation}
where $W_i$ is the waiting time in $M_2$ and $S_i$ is the service time at $C_2$. The waiting time $W_i$ is related to the system time of the $(i-1)$-th packet, $T_{i-1}$, and the inter-arrival time $Y_i$. In particular,
If $T_{i-1}>Y_i$, i.e., packet $i$ arrives at $M_2$ while the $(i-1)$-th packet is still waiting in queue or is under service, we have $W_i = T_{i-1}-Y_i$. Otherwise, $W_i=0$. Therefore, the waiting time of packet $i$ can be expressed as
\begin{equation}\label{equ:wait_time}
W_i =(T_{i-1}-Y_i)^+.
\end{equation}
From (\ref{equ:sys_time}), the term $E[T_{i}Y_{i-1}]$ can be given as
\begin{equation}\label{equ:ex_1}
\begin{split}
E[T_{i}Y_{i-1}]&=E[(W_{i}+S_{i})Y_{i-1}]\\
&=E[W_{i}Y_{i-1}]+E[S_{i}Y_{i-1}].
\end{split}
\end{equation}
According to (\ref{equ:sys_time}) and (\ref{equ:wait_time}), we can obtain the term $W_{i}$,
\begin{equation}\label{equ:}
\begin{split}
W_{i}& = (T_{i-1}-Y_{i})^+=(W_{i-1}+S_{i-1}-Y_{i})^+\\
 &= ((T_{i-2}-Y_{i-1})^{+}+S_{i-1}-Y_{i})^+.
 \end{split}
\end{equation}
We note that the system time $T_{i-2}$ relies on the waiting time and service time of packet $(i-2)$, hence is independent of $S_{i-1}$, $Y_{i}$ and $Y_{i-1}$. Moreover, the system will reach a steady state, thus the system times $T_{i}$ become stochastically identical, i.e., $T=^{st}T_i=^{st}T_{i-1}=^{st}T_{i-2}$.
The probability density function of the system time $T$ for the $M/M/1$ system is \cite{Papoulis2001Probability}
\begin{equation}\label{equ:}
f_{T}(t) = \mu_2(1-\rho)e^{-\mu_2(1-\rho)t},~~~ t\geq 0.
\end{equation}
The condition expected waiting time $W_{i}$ given $Y_{i-1} = y_i$ can be obtained as
\small
\begin{align}
%\label{equ:condi_1}
\begin{split}
&E[W_{i}|Y_{i-1} = y_i]=E[((T_{i-2}-y_{i})^{+}+S_{i-1}-Y_{i})^+|Y_{i-1} = y_i]\\
=&E[((T_{i-2}-y_{i})^{+}+S_{i-1}-Y_{i})^+]\\
=&\int_{0}^{\infty}f_{T}(t)\int_{0}^{\infty}f_{S}(s)\int_{0}^{\infty}f_{Y}(y)\left((t-y_i)^++s-y\right)^+dy dsdt,\\
=&\int_{0}^{y_i}f_{T}(t)\int_{0}^{\infty}f_{S}(s)\int_{0}^{s}f_{Y}(y)(s-y)~dydsdt \\ &+\int_{y_i}^{\infty}f_{T}(t)\int_{0}^{\infty}f_{S}(s)\int_{0}^{t-y_i+s}f_{Y}(y)(t-y_i+s-y)~dydsdt,\\
=&\frac{2\rho}{\mu_2(1+\rho)(1-\rho)}e^{-\mu_2(1-\rho)y_i}+\frac{\rho}{\mu_2(1+\rho)}.\label{equ:condi_solu}
\end{split}
\end{align}

Returning to (\ref{equ:ex_1}), the inter-arrival time $Y_{i-1}$ is independent of $S_{i}$, the service time of $C_2$ for the $i$-th packet, therefore, (\ref{equ:ex_1}) can be rewritten as
\begin{equation}\label{equ:ex_1_up}
E[T_{i}Y_{i-1}] =E[W_{i}Y_{i-1}]+E[S_{i}]E[Y_{i-1}],
\end{equation}
where $E[S_{i}]=\frac{1}{\mu_2}$. Further, utilizing the conditional expectation in (\ref{equ:condi_solu}), we can obtain the following equation
\begin{align}
&E[W_{i}Y_{i-1}]=\int_{0}^{\infty}y_{i}E[W_{i}|Y_{i-1}=y_i]f_{Y_i}(y_i)~dy_i\\
&\!=\!\int_{0}^{\!\infty}\!y_{i}\left(\frac{2\rho}{\mu_2(1+\rho)(1-\rho)\!}e^{-\mu_2(1-\rho)y_i\!}\!\!+\!\frac{\rho}{\!\mu_2(1\!+\!\rho)}\right)\mu_1e^{\!-\mu_1y_i\!}\!~dy_i\\
\label{equ:joint_1}&=\frac{2\rho^2-\rho+1}{\mu_2^2(1+\rho)(1-\rho)}.
\end{align}
Combining (\ref{equ:average_1}), (\ref{equ:exp_y}), (\ref{equ:exp_y_sq}),~(\ref{equ:ex_1_up}) and (\ref{equ:joint_1}), the average age (\ref{equ:average_up}) can be obtained.
%\subsection{comparison}
%\subsection{Baseline computing}

\subsection{Back to the computing models}
\subsubsection{AoI in local computing}
Comparing Fig. \ref{fig:subfig:a} with Fig. \ref{fig:jointfig}, local computing rate in Fig. \ref{fig:subfig:a} is equal to  $\mu_1$ in Fig. \ref{fig:jointfig}. Since the processed control signal size is much smaller than the original packet size, the time to transfer it to the destination node can be ignored. It is equivalent to the case  with infinite service rate for $C_2$ in Fig. \ref{fig:jointfig}. Denote local computing rate by $\mu_l$, the average AoI in local computing is
\begin{equation}\label{equ:average_up_local}
\Delta_{l} = \bar{\Delta}|_{\mu_1=\mu_l,\mu_2\rightarrow\infty}=\frac{2}{\mu_l}.
\end{equation}
\subsubsection{AoI in remote computing}In remote computing, the transceiver in Fig. \ref{fig:subfig:b} is equivalent to the server $C_1$ in Fig. \ref{fig:jointfig}, and the MEC server is equivalent to the server $C_2$. Denote the transmission rate by $\mu_t$ and the computing rate by $\mu_s$, the average AoI in remote computing is
\begin{equation}\label{equ:average_up_mec}
\Delta_{s} =\frac{1}{\mu_s}\left(\frac{\rho_s(2\rho_s^2-\rho_s+1)}{(1+\rho_s)(1-\rho_s)}+\frac{2}{\rho_s}+1\right),
\end{equation}
where $\rho_{s} = \mu_t/\mu_s$ is the ratio between transmission rate and computing rate.

\begin{figure}
\centering
    \includegraphics[width=3.5in]{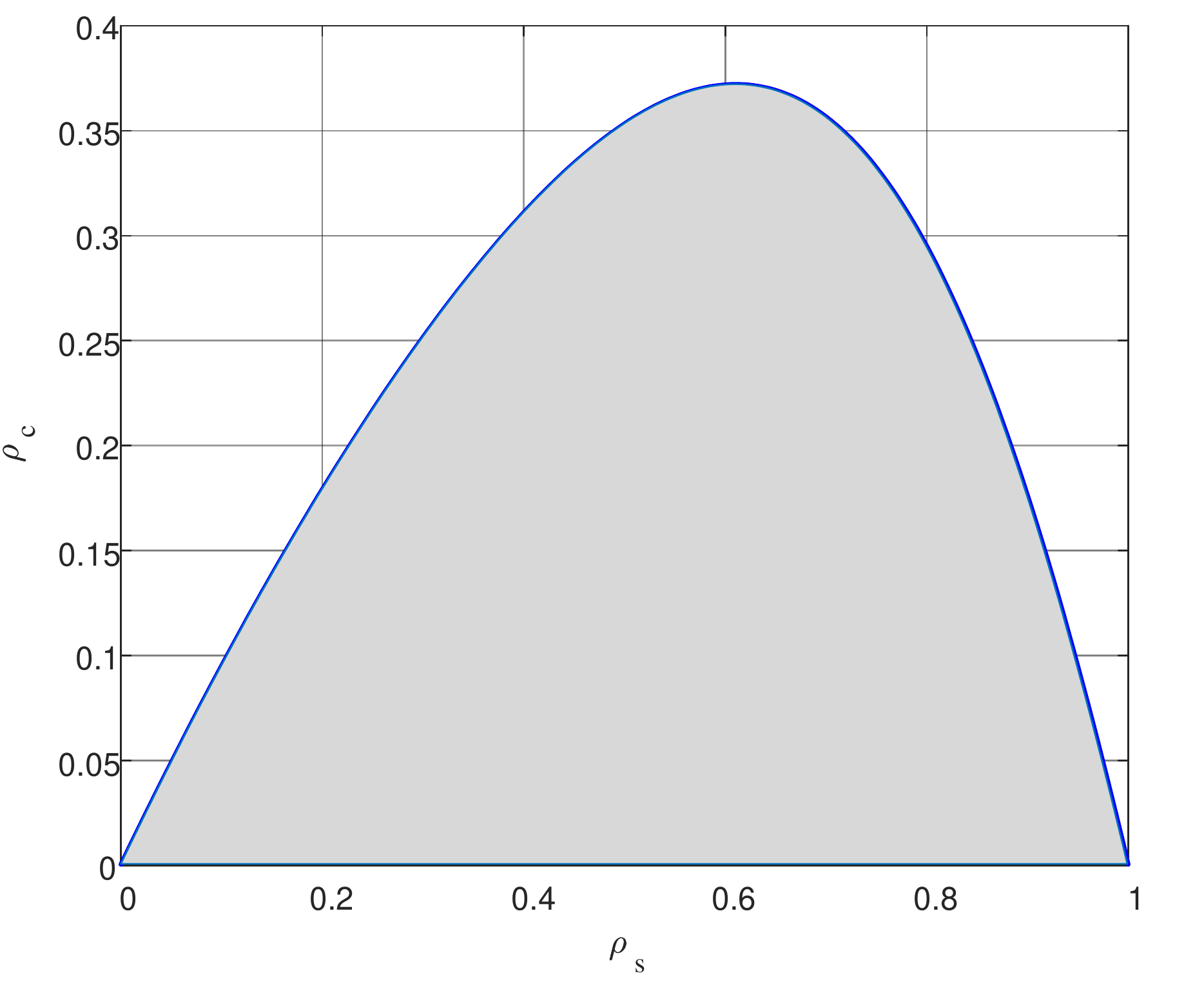}
\caption{Graphical representation of the regions where remote computing outperforms local computing  (shaded area). The blue curve corresponds to $\Delta_l=\Delta_s$.} \label{fig:com}
\end{figure}

With the above results, we can have a quick observation about when remote computing outperforms local computing. Let $\rho_c =\mu_l/\mu_s$, and depict the curve for $\Delta_l = \Delta_s$ as in Fig. \ref{fig:com}, we can find that in the region below the curve, i.e., the shaded area in the figure, remote computing can achieve smaller average AoI than local computing. As shown in the figure, when $\rho_s$ is close to 0 or 1, remote computing outperforms local computing only when $\rho_c$ is small. That is, for either small or large value of transmission rate, the local computing is better even with a small value of local computing rate. In other cases, remote computing is better in wider ranges. When $\rho_s \approx 0.61$, remote computing outperforms local computing even for $\rho_c \approx 0.37$. Therefore, there exists an optimal transmission rate so that remote computing is better for a largest range.

\section{Numerical Analysis}
In this section, we study the influence of parameters in MEC system on the average AoI, including packet size, required number of CPU cycles, average data rate and computing capacity of MEC server.
\subsection{Preliminary analysis}
 We use a pair $(l,c)$ to characterize the status update packet, where $l$ is the input size of the packet and $c$ indicates the required number of CPU cycles to compute the packet. Assume that all status update packets are of identical pair. The transmission time is related to the size of the transmitted data and the data rate. The computing time is associated with the required number of CPU cycles and computing capacity. Denote $f_1$ and $f_s$ as average local computing capacity and average computing capacity of MEC server allocated to the status update, respectively. Denote $R$ as the average data rate of the channel. Then the service rates $\mu_l$, $\mu_t$, $\mu_s$ can be expressed as
 \begin{equation}\label{equ:mul}
\mu_{l} = f(c,f_l) = \frac{f_l}{c},
\end{equation}
 \begin{equation}\label{equ:mu4}
\mu_{t} = f(l,R) = \frac{R}{l},
\end{equation}
 \begin{equation}\label{equ:mus}
\mu_{s} = f(c,f_s) = \frac{f_s}{c}.
\end{equation}
Accordingly, $\rho_s$ can be expressed as
\begin{equation}\label{equ:uti_mec}
\rho_s =\frac{Rc}{lf_s}.
\end{equation}
Intuitively, if the local computing capacity is large enough, the AoI in local computing can be smaller than that in remote computing. On the other hand, if the computing capacity of the MEC server is far superior to the local computing, the time saved by remote computing outweighs the additional time consumed for data transmission, therefore, the remote computing is better than the local computing. Similarly, if the data rate is large enough, the reduced computation time of remote computing can compensate for the time it takes to transmit the original packet. Correspondingly, if the packet size is small, while the amount of required CPU cycles is large, it should be transmitted to the MEC server. Therefore, the AoI between the local computing and the remote computing needs to comprehensively consider the packet size, required number of CPU cycles, local computing capacity, MEC server computing capacity, and data rate of the channel. Next, we will study the impact of these parameters by numerical results.

\subsection{Numerical results}

\begin{figure}
\centering
    \includegraphics[width=3.2in]{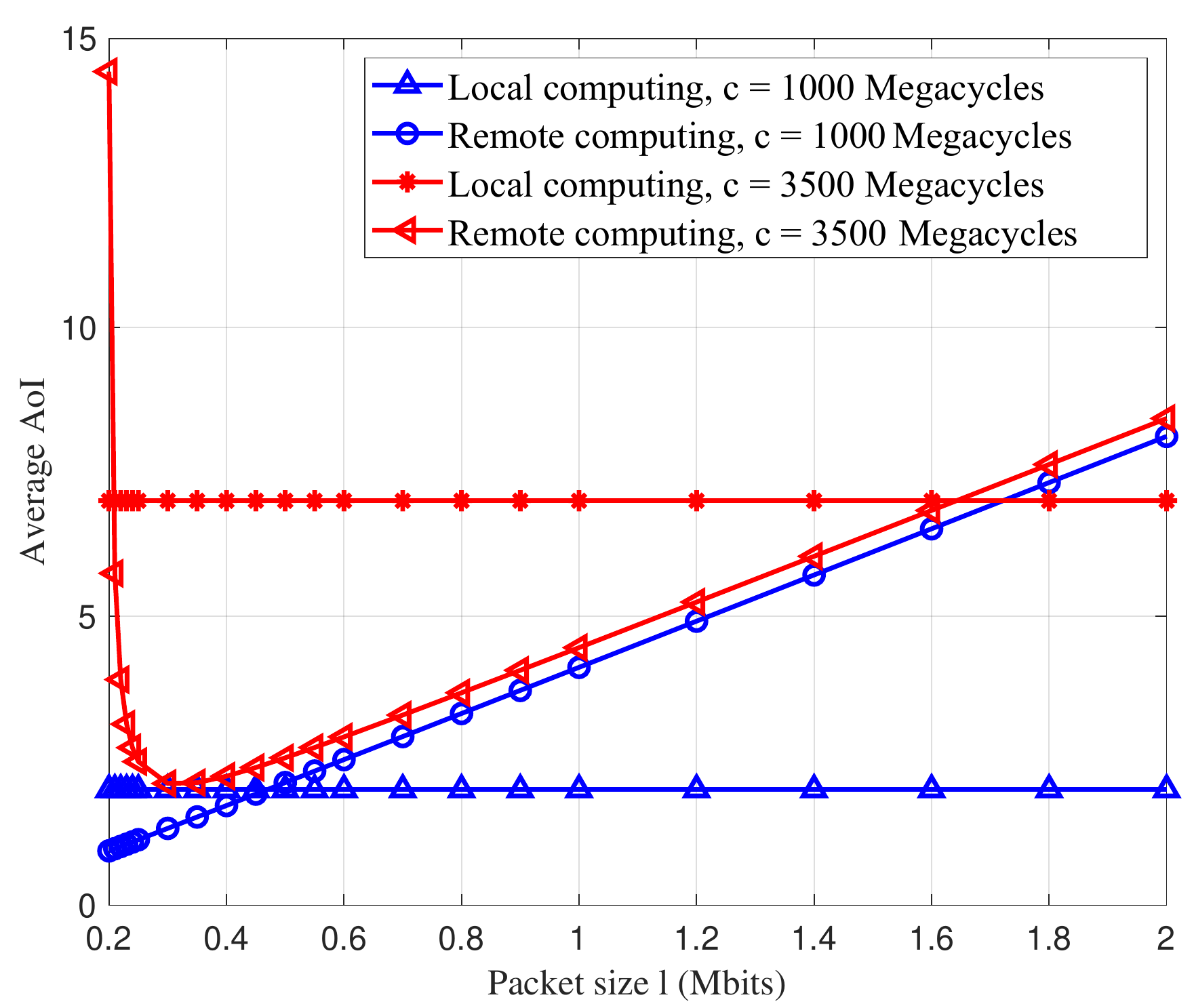}
\caption{AoI versus packet size} \label{fig:packet}
\end{figure}

\begin{figure}
\centering
    \includegraphics[width=3.4in]{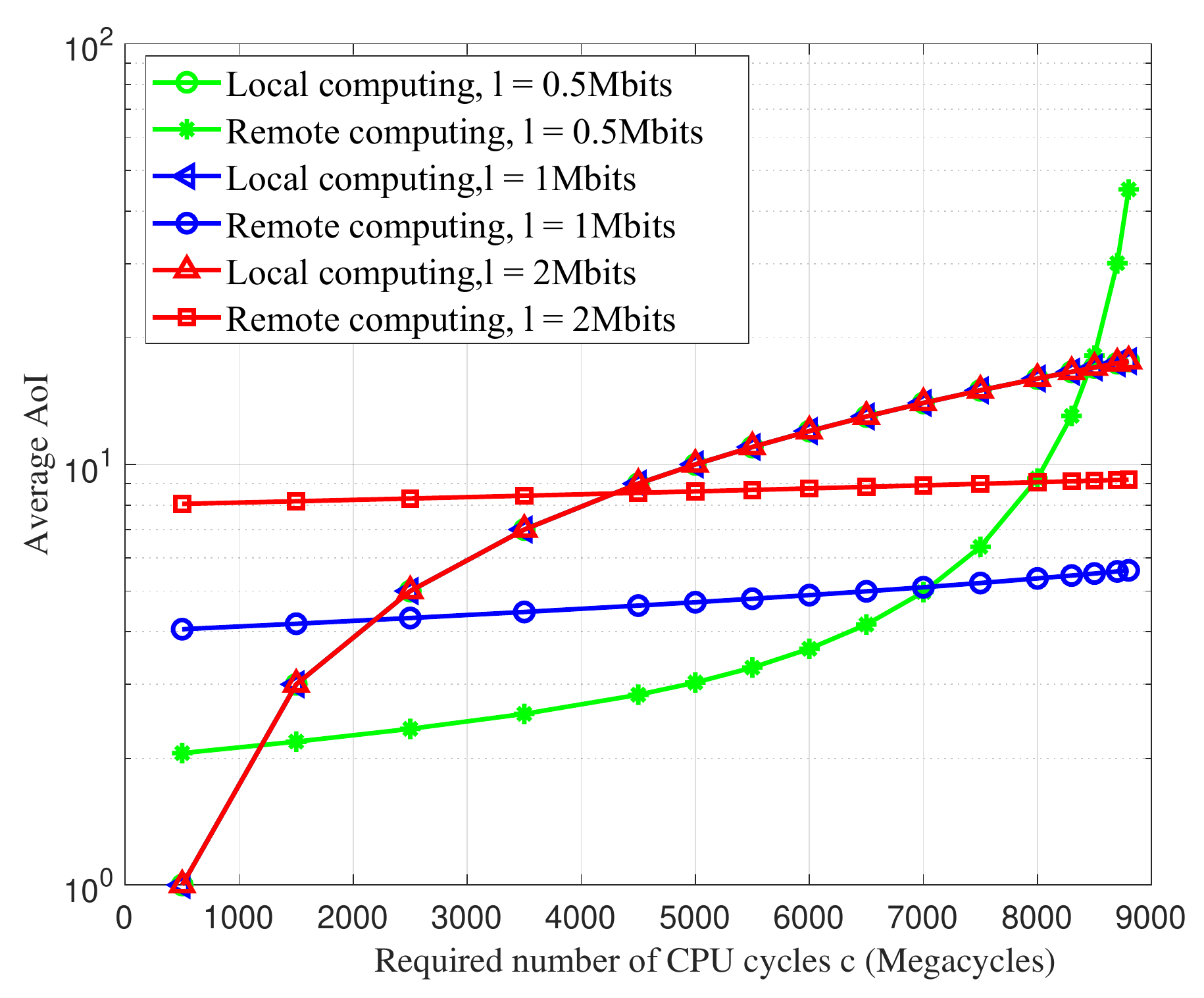}
\caption{AoI versus required number of CPU cycles} \label{fig:CPU}
\end{figure}

In the simulation of Fig. \ref{fig:packet} and \ref{fig:CPU}, we set the data rate as $R = 0.5$ Mbits/s, $f_l = 1$ GHz and $f_s = 9$ GHz.
Fig. \ref{fig:packet} shows the average AoI versus the packet size $l$ with different required number of CPU cycles in two schemes. As shown in Fig. \ref{fig:packet}, as the packet size increases, the AoI for local computing is constant, since the transmission time is ignored due to the small-sized packet after processing. For remote computing, because the required number of CPU cycles is fixed, the AoI is a function of the packet size, or a function of server utilization $\rho_s$. Analysing the AoI (\ref{equ:average_up_mec}) on the function of $\rho_s$, we know that when $\rho_s\approx0.61$, the AoI will reach the minimum; and the growth rate of AoI becomes very fast if $\rho_s> 0.9$; when $\rho_s$ tends to 1, AoI will be infinite, because there will be a lot of packets queued in the queue without being served. This is why there is a sudden drop and the minimum in remote computing with $c=3500$ Megacycles. It can be seen from Fig. \ref{fig:packet} that when $c= 1000$ Megacycles, the local computing and the remote computing lines cross at the point of $l\approx 0.47$ Mbits; while when $c=3500$ Megacycles, they cross at the point of $l\approx 1.64$ Mbits. This phenomenon implies that with appropriate small packet size, remote computing is superior to local computing while with large packet size, local computing is a better choice. Moreover, with the increase of packet size, the local computing performance will always surpass the remote one. The cross point is determined by the required number of CPU cycles.%Therefore, we know that the maximum throughput, i.e., server utilization $\rho_s=1$, can not make the AoI minimum.

Fig. \ref{fig:CPU} shows the average AoI versus the required number of CPU cycles $c$ with different packet sizes in two schemes. As shown in the figure, as the required number of CPU cycles increases, the AoI of local computing increases linearly. The size of the packet does not change the AoI for local computing. Therefore, the curves for local computing with three different packet sizes overlap. For remote computing, the AoI increases as the number of required CPU cycles increases due to the increased computation time. When the required number of CPU cycles is large, it is prone to the case where remote computing can achieve smaller AoI than the local computing, and it is easier to happen in smaller packet size. As shown in the figure, when $c\geq 7000$ Megacycles, the average AoI with $l=0.5$ Mbits sharply increases. This is because as $c$ increases, $\rho_s$ tends to 1, which results in the average AoI infinite.

\begin{figure}
\centering
    \includegraphics[width=3.4in]{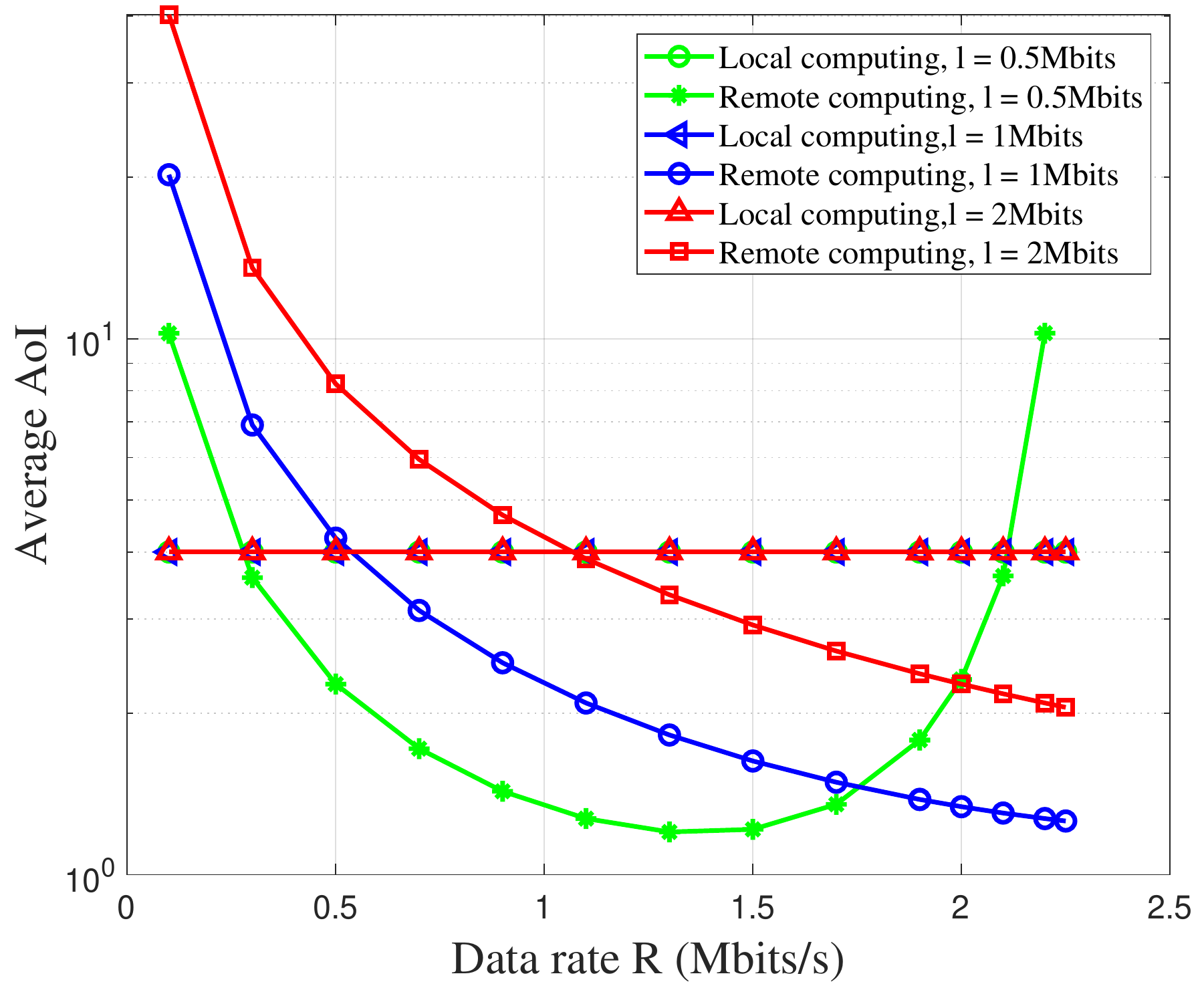}
\caption{AoI versus data rate} \label{fig:data}
\end{figure}

In the simulation of Fig. \ref{fig:data}, we set the required number of CPU cycles $c=2000$ Megacycles, $f_l = 1$ GHz and $f_s = 9$ GHz. As we can see from Fig.\ref{fig:data}, the three overlapped horizontal lines of local computing with different value of $l=0.5,1,2$ Mbits show that the average AoI of local computing is not affected by data rate $R$ and the packet size $l$.  For remote computing,  take $l=0.5$ Mbits as an example, we can see that the average AoI of remote computing will firstly decrease and then increase with the increasing of data rate. With the increasing of data rate, the average AoI of remote computing will be smaller than that of local computing. However, when the data rate continues to increase, the situation will be reversed. Therefore, there exists an optimal value of data rate to minimize the average AoI of remote computing.
\begin{figure}
\centering
    \includegraphics[width=3.4in]{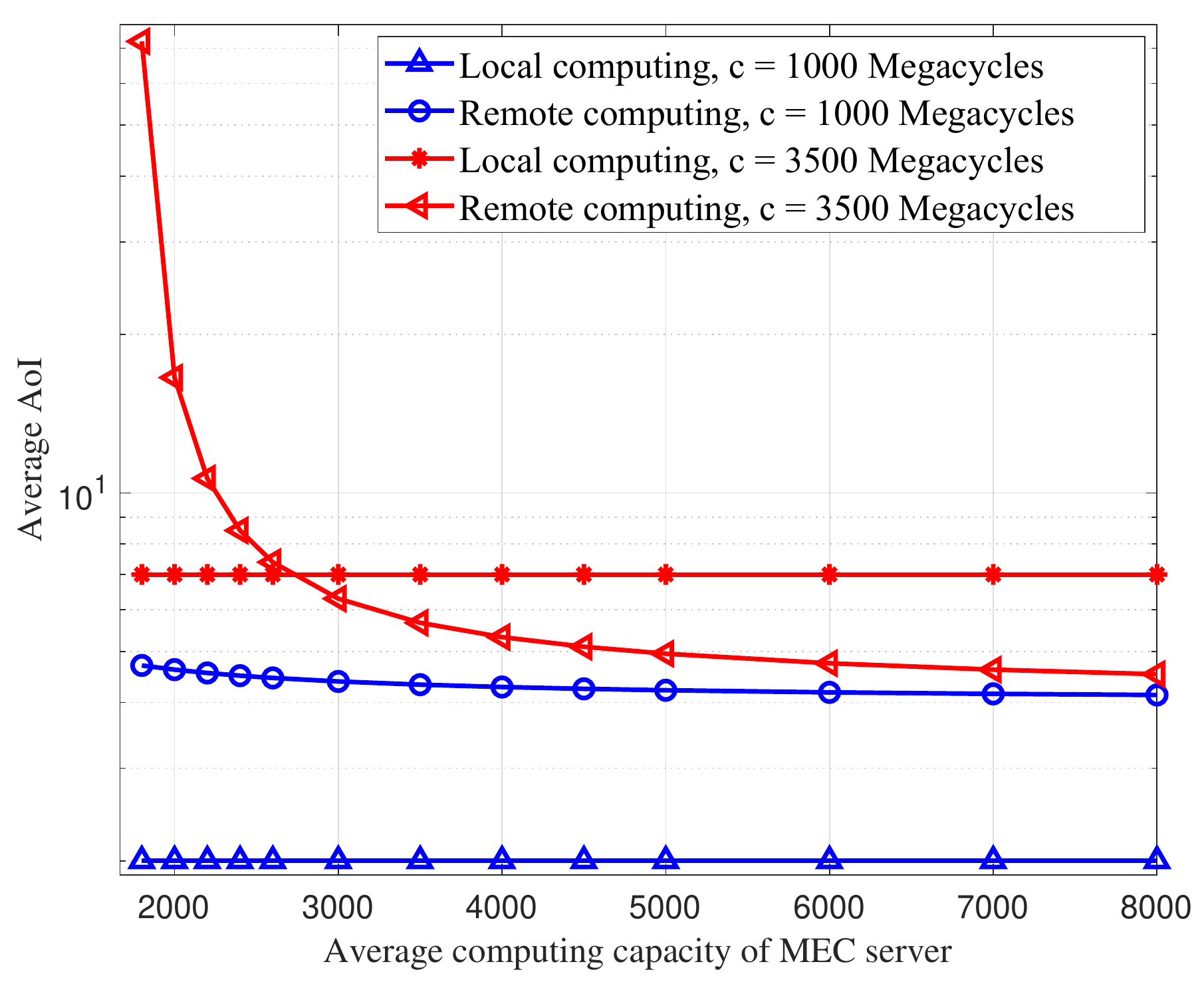}
\caption{AoI versus the average computing capacity of MEC server} \label{fig:computing}
\end{figure}
\balance

In the simulation of Fig. \ref{fig:computing}, we set the data rate $R=0.5$ Mbits/s,  local computing capacity $f_l = 1$ GHz and packet size $l = 1$ Mbits. When the number of required CPU cycles is small, for example $c=1000$ Megacycles, there will be no intersection between remote computing and local computing. While when the number of required CPU cycles is large, for example $c=3500$ Megacycles, the average AoI of remote computing will drop dramatically in the scope of lower average computing capacity of MEC server, and it will eventually converges to a stable level in the scope of higher value of average computing capacity of MEC server, the stable value is also the minimum average age, that is $\bar{\Delta}_{min}=\frac{2}{\mu_t}$. During the process of decreasing, there will be an intersection between remote computing and local computing, i.e., the average AoI of remote computing is smaller than local computing. Therefore, as the computing capacity of MEC server continues to increase, can remote computing be better than local computing, depending on the average local computing rate and transmission rate.
%\balance
\section{CONCLUSIONS}
In this paper, we consider the AoI for computation-intensive messages in MEC with two schemes, one is the local computing, and the other is remote computing. The closed-form average AoI for local computing and remote computing is derived, and the region where remote computing outperforms local computing is given. Numerical results showed that there exists an optimal transmission rate so that remote computing is better than local computing for a largest range. It is more likely that remote computing outperforms local computing at lager remote computing rate. As we can see, adopting MEC is crucial to obtain the optimal AoI for computation-intensive data. In the future works, it is worth extending the work to partial remote computing and multi source-destination pairs.
%\balance
\bibliographystyle{IEEEtran}
\bibliography{references_KQB0111}

\end{document}